\documentclass[manuscript]{aastex}

\usepackage{graphicx}
\usepackage{hyperref}
\usepackage{natbib}
\usepackage{textcomp}
\usepackage{gensymb}
\usepackage{url}
\usepackage{adjustbox}

\newcommand{\kms}{km s$^{-1}$}

\newcommand{\uJypb}{$\rm{\mu}$Jy beam$^{-1}$}
\newcommand{\mJypb}{mJy beam$^{-1}$}
\newcommand{\Jypb}{Jy beam$^{-1}$}

\shorttitle{Methanol Maser Survey}
\shortauthors{Hu et al.}

\begin{document}

\title{On the Relationship of UC~\ion{H}{2} Regions and Class II Methanol 
       Masers: I. Source Catalogs}

\author{B. Hu}
\affil{Purple Mountain Observatory, Chinese Academy of Science, 
       Nanjing 210008, China}
\affil{Max-Planck-Institut f\"ur Radioastronomie, 
       Bonn 53123, Germany}
\email{hubonju@gmail.com}

\author{K. M. Menten}
\affil{Max-Planck-Institut f\"ur Radioastronomie, 
       Bonn 53123, Germany}

\author{Y. Wu}
\affil{National Astronomical Observatory of Japan, 
       2-21-1 Osawa, Mitaka, Tokyo 181-8588, Japan}

\author{A. Bartkiewicz}
\affil{Centre for Astronomy, Faculty of Physics, Astronomy and Informatics, 
       Nicolaus Copernicus University, 
       Grudziadzka 5, 87-100 Torun, Poland}

\author{K. Rygl}
\affil{Osservatorio di Radio Astronomia (INAF-ORA), 
       Via P. Gobetti 101, 40129 Bologna, Italy}

\author{M. J. Reid}
\affil{Harvard-Smithsonian Center for Astrophysics, 
       Cambridge, Massachusetts 02138, USA}

\author{J. S. Urquhart}
\affil{School of Physical Sciences, University of Kent, 
       Ingram Building, Canterbury, Kent CT2 7NH, UK}

\author{X. Zheng}
\affil{School of Astronomy and Space Science, Nanjing University, 
       Nanjing 210008, China}

\begin{abstract}
  We conducted VLA C-configuration observations to measure positions and 
  luminosities of Galactic Class II 6.7~GHz methanol masers and their 
  associated ultra-compact \ion{H}{2} regions.  The spectral resolution was 
  3.90625 kHz and the continuum sensitivity reached 45 \uJypb.  We mapped 372 
  methanol masers with peak flux densities of more than 2 Jy selected from 
  the literature, 367 of them were detected.  
  Absolute positions have nominal uncertainties of 0.3\arcsec.  
  In this first paper on the data analysis, we present three catalogs,
  the first gives information on the strongest feature of 367 methanol maser 
  sources, and the second on all detected maser spots. The third catalog 
  present derived data of the 279 radio continuum sources found in the vicinity
  of maser sources. Among them, 140 show evidence of physical association with
  maser sources.  Our catalogs list properties including distance, flux 
  density, radial velocity and the distribution of masers on the 
  Galactic plane is then provided as well. We found no significant 
  relationship between luminosities of masers and their associated radio 
  continuum counterparts.
\end{abstract}

\keywords{catalog --- ISM: molecules --- masers --- stars: formation}

\section{Introduction}
\label{sec:introduction}

  % Methanol maser traces the high mass star forming region
  Although massive stars make up only a few percent of the stellar population 
  in the Milky Way, they play a central role in many astrophysical processes, 
  such as shaping the interstellar medium, regulating star formation and 
  ultimately governing the evolution of their host 
  galaxy~\citep{2005IAUS..227....3K}. 
  However, to date, understanding massive star formation remains a challenge,
  because of numerous observational problem.  Nevertheless, investigations 
  of the natal environment of massive stars provide rich
  clues for a better understanding of its physical and chemical properties
  and the star formation process itself.
  
  Massive star forming regions (MSFRs) offer a variety of tracers that can
  be detected in different wavebands, including massive dense 
  cores~\citep{2007ARA&A..45..481Z}, infrared dark 
  clouds~\citep{2010ApJ...723L...7K}, mid-infrared 
  sources~\citep{2013ApJS..205....1P}, Extended Green 
  Objects~\citep[EGOs,][]{2008AJ....136.2391C}, molecular 
  outflows~\citep{2002A&A...383..892B}, 
  masers~\citep[most common species are hydroxyl, methanol and 
  water;][]{2014IJAA....4..571F}, compact and ultra-compact~(UC) \ion{H}{2} 
  regions~\citep{2002ARA&A..40...27C} 
  and others.  Understanding what these signposts tell us may help to
  construct an evolutionary sequence for young massive stars. 

  Methanol masers at 6.7~GHz were discovered by \citet{1991ApJ...380L..75M}; 
  they, along with 22.2 GHz H$_2$O masers, are 
  the brightest and most widespread maser species in massive star forming 
  regions~\citep[e.g.][]{1995MNRAS.272...96C,2003A&A...403.1095M,
  2008A&A...485..729X}.  These masers may be
  one of the earliest and ubiquitous tracers of the massive star 
  formation process~\citep{2015MNRAS.446.3461U}. Most 6.7~GHz methanol masers 
  have been observed with a variety of telescopes and with differing 
  resolutions.  For example, \citet{2011ApJ...730...55P} observed 57
  methanol masers at 6.7~GHz with the Multi-Element Radio-Linked Interferometer 
  Network~(MERLIN) with an angular resolution of 60~mas.   At this resolution
  they found a close correspondence between the masers and mid-infrared sources 
  and claimed their results to support the theoretical model that 
  these masers are pumped by infrared dust emission in the vicinity 
  of massive proto-stars.   

  %However, the lack of a higher 
  %resolution methanol maser catalog, currently limits direct association of
  %the masers with an embedded young star.
  
  % cornish survey
  Compact and ultra-compact~(UC) \ion{H}{2} regions are clear indicators
  of sites of massive stars~\citep{2005dmgp.book..203H}, whose ionizing 
  photons are provided by embedded OB stars in the later stages of massive 
  star formation~\citep{2002ARA&A..40...27C}. The thermal bremstahlung 
  radiation from the ionized gas makes \ion{H}{2} regions directly 
  observable in 
  radio continuum bands. In recent decades, a number of radio survey have 
  identified many hundreds of UC \ion{H}{2} regions. For example, 
  \citet{2009A&A...501..539U} observed 659 MSFR candidates with the VLA at 
  4.9~GHz and an angular 
  resolution of 1.5", achieving  RMS noise levels of $\approx0.2$~\mJypb\ and 
  detected 391 compact or UC~\ion{H}{2} regions.
  Many observations~\citep[e.g.][]{2009PASA...26..454C,2010MNRAS.407.2599C,
  2010A&A...517A..78S} have
  established an association of these \ion{H}{2} regions with 6.7~GHz methanol 
  masers. Since compact and UC~\ion{H}{2}
  regions present a reliable snapshot of MSFRs withing the last $\sim10^5$
  years~\citep{2011MNRAS.416..972D,2011ApJ...730L..33M}, a census on this later 
  phase of MSFRs 
  can help us to better understand the evolutionary sequence of massive
  star formation and their association with different tracers.   
  
  % The motivation of this paper
  In order to study the relationship of 6.7~GHz methanol masers to 
  compact and UC \ion{H}{2} regions,
  we conducted a large survey to simultaneously observe the methanol maser line 
  and C-band continuum emission with the Karl G. Jansky Very Large Array~(VLA)
  in its C-configuration.  We mapped 372 class II methanol maser sources at 
  6.7~GHz, selected from literature with peak fluxes $>2$~Jy.  
  Absolute positions were established based on phase-referenced observations
  to astrometric calibrators.  Our observations also served as precursor 
  observations for the Bar and Spiral Structure Legacy~(BeSSeL) 
  Survey~\citep{2011AN....332..461B} to provide high
  accuracy position of maser sources necessary for VLBA observations. 

  Our uniform and high angular resolution survey includes the vast
  majority of known 6.7~GHz methanol masers and their nearby compact or 
  UC~\ion{H}{2} regions. 
  The survey produced three catalogs: (1) a methanol maser catalog containing
  accurate positions, radial velocities and flux densities of the brightest  
  spots; (2) a catalog of maser spectra and spot maps; (3) a catalog of 
  \ion{H}{2} regions toward the methanol masers. 

  Compact \ion{H}{2} (referring to both compact and UC \ion{H}{2}) regions 
  and methanol masers offer a unique opportunity to peer into 
  the deep interior of massive star forming regions at high angular resolution
  since at centimeter wavelength their natal clouds are optically thin,
  and the large database should allow statistically meaningful estimates of 
  many characteristics of 6.7~GHz methanol masers and identified \ion{H}{2} 
  region counterparts.  In combination with other signposts of massive star 
  formations, the catalogs may provide important clues to solve several key 
  issues related to massive star formation, such as whether 6.7~GHz methanol 
  masers are located in accretion disks or in outflow walls (or both), 
  when in the process of star formation these two signposts occur, 
  how MSFRs evolve with time. Some prominent class II methanol masers are 
  found in the dense molecular material associated UC \ion{H}{2} regions, 
  W3OH and NGC~7538 being prominent examples~\citep{1988ApJ...333L..83M}. 
  However, most have 
  no radio continuum emission at the, say, few mJy level, that could be 
  easily detected with the original VLA or the 
  ATCA~\citep[e.g.][]{1998MNRAS.301..640W, 2001A&A...369..278M}.  
  This leaves open the question of 
  whether weaker radio emission could be associated with masers at the level 
  that could be expected from younger, hyper compact \ion{H}{2} regions.
  
  In Section~\ref{sec:Observ} we describe the sample and the design of the
  observations.  The data analysis is described in Section~\ref{sec:Reduc}.
  The three catalogs mentioned 
  above are presented in Section~\ref{sec:Produc}. In Section~\ref{sec:Prop}, 
  statistical properties of the 6.7~GHz methanol masers and their associated 
  compact \ion{H}{2} regions are discussed. A brief summary is given 
  in Section~\ref{sec:Summary}.   This paper is the first of a series 
  investigating massive star formation and its 
  evolutionary sequence.

\section{Observation Design}
\label{sec:Observ}

  \subsection{Sample}
  \label{sec:Observ_Sample}
    % source selection
    We selected our sample of 6.7~GHz methanol masers based on the 
    following criteria:
    \begin{enumerate}
      \item Sources should have declination higher than $-30\degr$ to
            ensure the visibility from VLA site.
      \item Since subsequent VLBA observations for the BeSSeL Survey require 
            high SNR for parallax measurements, we required a peak flux 
            density greater than 2~Jy.
    \end{enumerate}

    Taking these constraints into account, 372 unique targets were selected 
    from the following methanol maser surveys: the Methanol Multi-Beam catalog
    (MMB)~\citep{2010MNRAS.407.2599C, 2010MNRAS.409..913G,2012MNRAS.420.3108G},
    Arecibo Methanol Maser Galactic Plane Survey 
    (AMGPS)~\citep{2011ApJ...730...55P}, the Torun catalog of 6.7~GHz methanol 
    masers \citep{2012AN....333..634S}, and other individual
    observations of known 6.7~GHz methanol masers or massive star
    forming regions~\citep{1995MNRAS.272...96C, 1997MNRAS.291..261W, 
    1998MNRAS.301..640W, 2009PASA...26..454C, 2008A&A...485..729X}. 
    % completeness statement
    \citet{2009MNRAS.392..783G} state that the
    MMB survey is complete to $\sim$80\% at 0.8~Jy and $\approx100$\% at 1~Jy. 
    At the time we started our observation, MMB had published 
    sources in Galactic longitude range from 186\degr\ to 20\degr\ , and 
    Galactic latitude range from -2\degr\ to 2\degr.  However, most of the MMB 
    Survey covers the Southern Galactic plane and is not visible to the VLA. 
    Our observations 
    included MMB sources in Galactic longitude range from 357\degr\ to 
    20\degr. With a flux cutoff at 2~Jy, we believe the sample in this area 
    is complete.  The AMGPS methanol maser survey uniformly sampled from 
    35.2\degr\ $\leq l \leq$ 53.7\degr\ with $|b| \leq$ 0.41\degr\ and was 
    complete to 0.27~Jy~\citep{2007ApJ...656..255P}. Thus, our catalog is also
    complete to 2~Jy in the AMGPS area.  The completeness of the Torun catalog
    varies from Galactic longitude, however, \citet{2012AN....333..634S} conclude
    that all maser sources brighter than 7.5~Jy in the peak were included.
    So our survey is not complete to 2~Jy for these sources, nor from 
    sources that were from targeted observation towards individual objects.
    Most of the masers are located in the first and second quadrant of the 
    Galactic plane.

  \subsection{Observation Design}
  \label{subsec:Observ_Observ}
    % observation strategy
    The observations were conducted with the VLA in C-configuration using five 
    sessions from 2012 February 28th to April 16th. At the beginning of each 
    observation, a primary flux calibrator\footnote{Calibrators used are 3C147, 
    3C286 and 3C48. Their assumed flux densities are determined by the radio 
    flux calibrators model files delivered with the Common Astronomy Software 
    Applications (CASA).} was observed to establish amplitude calibration, 
    followed by observation alternating between phase calibrator and target
    sources.

    We adopted complex gain calibrators from the catalogs of 
    \citet{2011ApJS..194...25I} and the first realization of the International 
    Celestial Reference Frame~\citep[ICRF1,][]{1997AAS...191.1613M}.  Phase 
    calibrators were chosen to be within 5\degr\ of the targets and were 
    observed at the beginning and end of each 
    block.  So a typical cycle consists of a scan of phase calibrator,
    followed by a group of snapshots of nearby targets, then a revisit of
    the phase calibrator.

    % backend setup
%    We designated a specific back-end setup, simultaneously considering of both
%    high spectral resolution for maser line observation and large band width 
%    coverage for continuum detection. 
    Spectral line data used 2048 channels across 8~MHz, yielding channel
    spacing of 3.90625~kHz at the central frequency of 6.6685192~GHz,
    yielding velocity resolution of 0.176~\kms. 
    The continuum observations employed two 1~GHz sub-bands 
    from 4.9840~GHz to 6.0080~GHz (hereafter, the low band) and from 
    6.6245~GHz to 7.6485~GHz (hereafter, the high band) and each sub-band was
    divided in to 16 channels.  Snapshot observations of the maser sources 
    achieved about 20~seconds on-source and reached typical noise levels of 
    45~\uJypb\ for the continuum maps and 33~\mJypb\ per channel for the
    spectral line maps.  Technical details of the observations are given 
    in Table~\ref{tab:ObservTechDetails}.

    %%table of technical details
    \begin{deluxetable}{ll}
      \tablecolumns{2}
      \tablewidth{0pc}
      \tablecaption{A summary of technical details of the observation
                    \label{tab:ObservTechDetails}}
      \tablehead{}
      \startdata
      Dates of observations   & 2012 Feb. 28\\
                              & 2012 Mar. 04\\
                              & 2012 Mar. 07\\
                              & 2012 Mar. 18\\
                              & 2012 Apr. 16\\
      Configuration           & C-configuration \\
      Antennae                & 27 \\
      Digital Sampler         & 8-bits \\
      Average On Source Time  & 20 s \\
      Synthesized Beam        & $3\arcsec \times 6\arcsec$ \\
      field of view           & $\sim 10'\times 10'$ \\
      Frequency Coverage      & Line: 6664.5 to 6672.5~MHz\\
                              & Cont.: 4984.0 to 6008.0~MHz \\
                              & and 6.6245 to 7.6485~MHz \\
      \enddata
    \end{deluxetable}

\section{Data Reduction}
\label{sec:Reduc}

  \subsection{Methanol Maser Line}
  \label{subsec:Reduc_Line}
    Maser data were calibrated through a standard spectral-line data 
    reduction procedure in Astronomical Image Processing System (AIPS).
    In brief, we first flagged suspicious identified entries by eye, then 
    determined the flux density scale based on the primary flux calibrators. 
    Complex gains of secondary calibrators were then calculated and  
    interpolated to the maser sources.
    We extracted a scalar averaged spectrum from the visibility data of each
    target in order to find the peak emission channel.  Then we made an image
    from that channel's data covering $5.12'\times 5.12'$ on the sky, and 
    a detection was confirmed by requiring the peak brightness to exceed
    five times the RMS noise level of the image.  We then generated 
    a spectroscopic image cube with typical CLEAN restoring beam size of 
    $3\arcsec \times 6\arcsec$. 
    The positions of the maser spots were derived from these cubes..

    In a small number of cases (33), 
    we failed to detect maser emission, in spite of the fact
    that maser sources were reported there in the literature.
    Positions for the centers of these fields are given in 
    Table~\ref{tab:Non-detection}. In all case we made maps of 
    5$\times$5~arc min, we suspect that many of these non-detection results
    from previous catalogs having identified telescope side-lobes from a 
    strong source at a position far removed from the true source.
    However, for some of the weaker sources, it is possible that time 
    variability could explain our lack of detection ($< 5\sigma$ i.e. 
    about 170~mJy).  Repeated observations with longer integration might 
    reveal if variability explains out non-detections.

    %% non-detection table
    \begin{deluxetable}{llll}
      \tablecolumns{4}
      \tablewidth{0pc}
      \tablecaption{Non-detection sources\label{tab:Non-detection}}
      \tablehead{
      Source Name    & R.A.           & Dec.            & Reference \\
                     & (hh:mm:ss.sss) & (ddd:mm:ss.sss) &           
      }
      \startdata
      G000.69$-$0.04 &  17:47:24.74   &  $-$28:21:43.6    & C10 \\
      G000.71$-$0.03 &  17:47:25      &  $-$28:21:00      & W97 \\
      G002.62$+$0.14 &  17:51:12      &  $-$26:37:00      & W97 \\
      G010.00$-$0.03 &  18:07:52      &  $-$20:18:00      & W97 \\
      G010.20$-$0.35 &  18:09:28.43   &  $-$20:16:42.5    & G10 \\
      G012.86$-$0.27 &  18:14:36.42   &  $-$17:54:50.2    & S14 \\
      G016.60$-$0.04 &  18:21:09      &  $-$14:31:00      & W97 \\
      G016.88$-$2.15 &  18:29:25      &  $-$15:15:00      & W97 \\
      G018.67$+$0.03 &  18:24:53.78   &  $-$12:39:20.40   & G10 \\
      G019.37$-$0.02 &  18:26:24      &  $-$12:03:00      & W97 \\
      G023.09$-$0.39 &  18:34:45.71   &  $-$08:55:55.2    & S14 \\
      G023.45$-$0.18 &  18:34:39      &  $-$08:31:00      & W97 \\
      G023.46$-$0.16 &  18:34:35.95   &  $-$08:29:32.1    & S14 \\
      G023.71$-$0.19 &  18:35:11      &  $-$08:17:00      & W97 \\
      G025.23$+$0.29 &  18:36:17.10   &  $-$06:43:11.8    & S14 \\
      G025.66$+$1.06 &  18:34:20      &  $-$05:59:00      & W97 \\
      G025.70$+$0.03 &  18:38:05      &  $-$06:25:00      & W97 \\
      G025.79$-$0.14 &  18:38:52      &  $-$06:25:00      & W97 \\
      G025.80$-$0.16 &  18:38:56.40   &  $-$06:24:52.7    & SZ12\\
      G025.81$-$0.04 &  18:38:32.02   &  $-$06:21:28.6    & S14 \\
      G027.56$+$0.08 &  18:41:20.26   &  $-$04:44:25.0    & S14 \\
      G028.31$-$0.37 &  18:44:19      &  $-$04:17:00      & W97 \\
      G029.91$-$0.05 &  18:46:05.9    &  $-$02:42:27      & X08 \\
      G030.52$+$0.10 &  18:46:41.37   &  $-$02:06:07.6    & SZ12\\
      G030.79$+$0.20 &  18:46:48      &  $-$01:48:46      & W98 \\
      G030.82$+$0.27 &  18:46:37      &  $-$01:45:00      & W97 \\
      G031.28$+$0.07 &  18:48:10.02   &  $-$01:26:31.6    & SZ12\\
      G031.42$+$0.32 &  18:47:33      &  $-$01:12:00      & W97 \\
      G032.77$-$0.06 &  18:51:21.60   &  $-$00:10:04.9    & S14 \\
      G049.27$-$0.34 &  19:23:06.95   &  $+$14:20:10.9    & S14 \\
      G049.67$-$0.46 &  19:24:19.60   &  $+$14:38:03.9    & SZ12\\
      G189.77$+$0.34 &  06:08:32.40   &  $+$20:39:25.3    & SZ12\\
      G358.39$-$0.48 &  17:43:37.83   &  $-$30:33:51.1    & C10 \\    
      \enddata
      \tablecomments
        {
        Source names and coordinations are taken directly from the references. 
        J2000 coordination are presented in sexagesimal in 
        corresponding units. Accuracy of the coordination varies by 
        different references.  Fluxes are the peak values that 
        taken directly from the references. Reference abbreviations are:
        C10~\citep{2010MNRAS.407.2599C}; G10~\citep{2010MNRAS.409..913G};
        S14~\citep{2014A&A...563A.130S}; SZ12~\citep{2012AN....333..634S}; 
        W97~\citep{1997MNRAS.291..261W}; X09~\citep{2009A&A...507.1117X}. 
        }
    \end{deluxetable}

    We improved the calibration of the maser sources, using the brightest 
    maser spot to self-calibrate phase and amplitude and then applying these
    corrections to all channels.  For a maser spot with peak flux density 
    $>$3~Jy, 
    self calibration improved the dynamic range of the maser image cube.
    To extract maser spots from image cubes, we ran the SAD task on 
    each channel, with a dynamic range cutoff at 10\% of the image peak. 
    In some cases, more than one maser source appeared in a field.  When
    the separations exceeded $30\arcsec$, we labeled them as separate maser 
    sources.  The spectrum for each maser source was generated from the SAD 
    output results and has a velocity resolution of 0.176~\kms.

  \subsection{Continuum}
  \label{subsec:Reduc_Cont}

    We processed the continuum data using the Common Astronomy Software 
    Applications~(CASA).  Our calibration procedure can be summarized as 
    follows: after inspecting and flagging the data set, the flux density scale 
    was set to the primary calibrator; bandpass and antenna based delays were 
    then calculated for the primary calibrator and applied to the secondary 
    calibrators and targets; we then finish the calibration by interpolating
    the amplitude and phase correction from secondary calibrators to targets; 
    and after some additional flagging, images were produced with pixel size of
    0.3\arcsec.  Most continuum images are of $1024\times1024$ pixels, for some
    cases larger images are produced to include nearby by strong sources to
    suppress the interference from them.
    Since there is considerable radio frequency interference at C-band at 
    the VLA site, we had to manually flag a considerable amount of data.  
    In some cases this significantly increased the noise level in the final 
    image.  In addition, the flagging, together with snapshot observing, 
    produced limited (u,v)-coverage, which led to large side-lobes in 
    the final images, especially for sources at low Declination.

\begin{figure}
      \resizebox{\hsize}{!}{\includegraphics{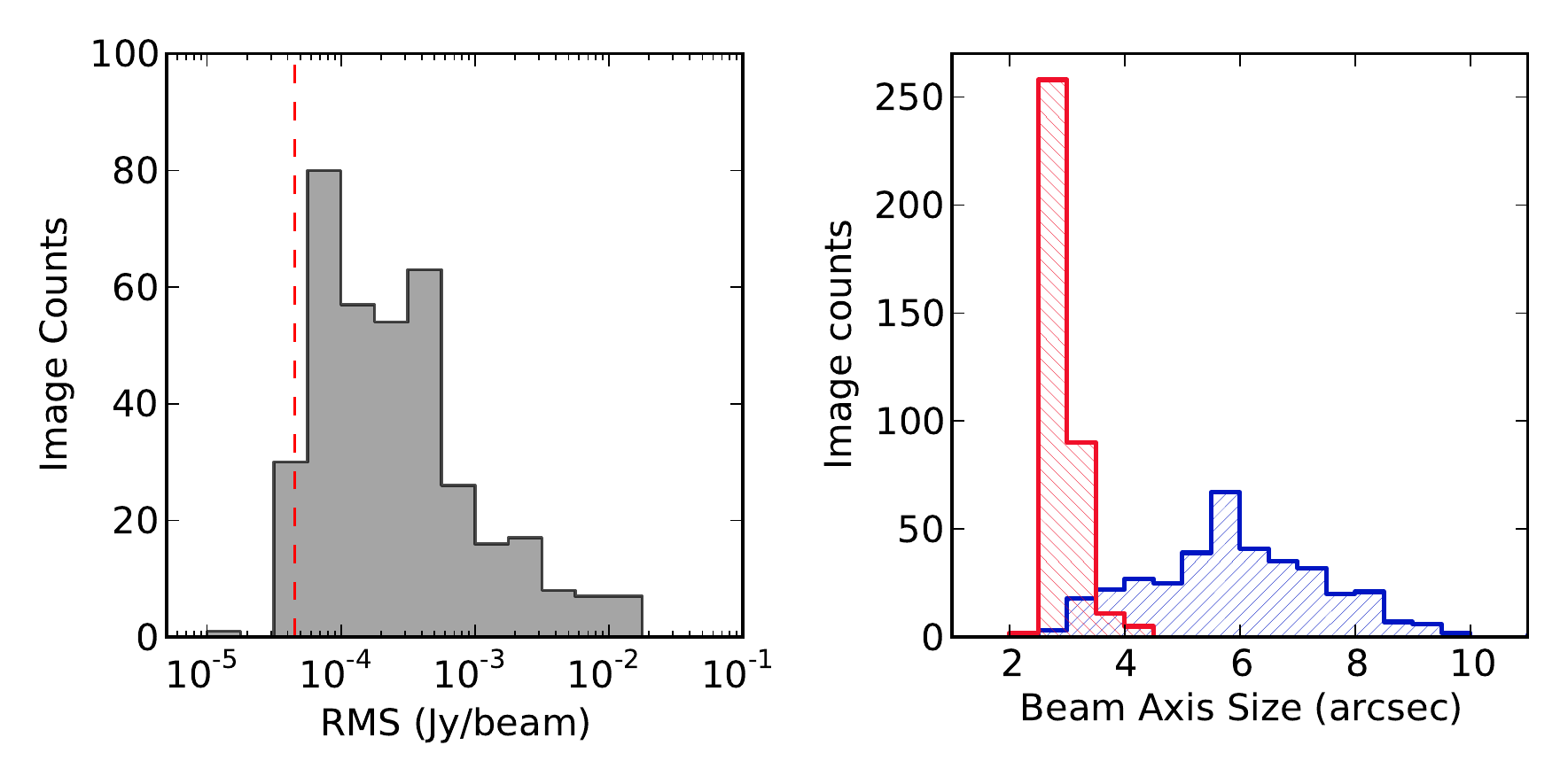}}
      \caption
        {
        Statistical properties of the continuum images. {\it Left panel:}
        the gray histogram indicates the distribution of RMS map noise,
        and the vertical red dashed line shows the expected thermal noise limit.
        {\it Right panel:} the distribution of the major ({\it blue}) and minor 
        {\it red} axis FWHM sizes of the Gaussian restoring beams.   
        A typical beam size for the survey is $6\arcsec \times 3\arcsec$.
        }
      \label{fig:ImgStat}
    \end{figure}

    The final maps were made with the CLEAN task in CASA, using the 
    ``multi-frequency synthesis'' algorithm. Three maps were produced for 
    each source, including those derived separately for the low- and high-band 
    data and a combined map.  We excluded channels near the 6.7 GHz maser line 
    when making these maps.  Typical noise levels for the combined map were 
    $\sim$50~\uJypb.  The left panel of Figure~\ref{fig:ImgStat} shows the
    distribution of RMS noise levels for the combined maps. The peak of
    the histogram is comparable to the theoretical noise level 
    ($\sim$45~\uJypb) for our observations as calculated by the VLA 
    Exposure Calculator\footnote{\url{https://obs.vla.nrao.edu/ect/}}. 
    For maps that have much greater noise than expected, most were affected 
    by side-lobes of nearby strong sources and complications from strong 
    diffuse background emission. 
    
    In order to extract positions of radio continuum sources in the fields,
    we used the SExtractor algorithm~\citep{1996A&AS..117..393B} to 
    automatically find and measure compact radio sources.  Although, 
    SExtractor was developed for source extraction in visible or infrared 
    images, it has been already successfully used for radio sources in the 
    past~\citep{2006A&A...453.1003T, 2013A&A...549A..45C, 2015MNRAS.446.3461U}. 
    The continuum sources identified by SExtractor were 
    filtered before inclusion in our catalog. We required the peak 
    flux to be above $3\sigma$ of the RMS noise of the integrated map, or above 
    $2\sigma$ on each of the 3 images (i.e., the integrated image, the 
    high-band image, and low-band image).   Under these constraints,
    349 continuum sources were identified.      
    For the interest of investigation of maser to continuum source 
    associations, we adopted only continuum source with the smallest 
    angular offset to the strongest maser spot in each maser source, 
    if multiple continuum sources appear in the same field. After all 
    279 sources are adopted, making up the continuum source catalog.

\section{Survey Products}
\label{sec:Produc}

  % maser catalogue
  \subsection{VLA 6.7~GHz Methanol Maser Catalog}
  \label{subsec:Produc_MaserCatalogue}
  
    Our VLA 6.7~GHz methanol maser catalog in Table~\ref{tab:Maser} includes 
    367 methanol maser sources, sorted by Galactic longitude, with 
    $\approx0.3\arcsec$ accurate 
    J2000 coordinates, peak ($S_{\rm{peak}}$) and integrated flux 
    ($S_{\rm{int}}$) flux densities, LSR velocities ($V_{\rm{lsr}}$) for 
    the strongest spots, the velocity ranges of masers ($\Delta V$), and 
    the estimated heliocentric distance and its probability 
    (see Section~\ref{subsec:Prop_Distance} for details).  
    The LSR velocity, and peak and 
    integrated flux densities were estimated by fitting two-dimensional 
    Gaussian brightness distributions of the reference spot.

    Positions of the brightest, compact maser spots were determined by 
    phase calibrator observations prior to any self-calibration. 
    The formal precision in determining position is directly 
    proportional to the synthesized beam size and inversely to the 
    signal-to-noise ratio (SNR)~\citep[see][Equation 1]{1988ApJ...330..809R}. 
    For the typical beam size $3\arcsec$ (E-W), $6\arcsec$ (N-S) and 
    SNR value of 50, in our observation, 
    the position precision is $\approx27$~mas in the E-W direction and
    $\approx54$~mas in the N-S direction. Thus we list the right ascension of
    maser spots to 3 decimals in second, and declination to 2 decimals in 
    arc~second in the catalog.  We've taken these values as the
    formal precision of maser spots position.  However, uncompensated 
    atmospheric delays likely make the random error much smaller than the 
    systematic error.  Based on experience with snap-shot imaging of OH masers 
    by \citet{2000ApJS..129..159A} and a comparison with maser positions 
    determined from VLBI observation (see Section~\ref{subsec:Prop_Positions} 
    for details), we adopt a realistic total position accuracy which is 
    $\pm0.3\arcsec$.

    %% maser catalogue
    \begin{deluxetable}{rrrrrrrrrr}
      \rotate
      \tablecolumns{10}
      \tablewidth{0pc}
      \tablecaption{6.7~GHz methanol maser catalog\label{tab:Maser}}
      \tablehead{
        %% Header
          \colhead{Source name}       %% Source Name  1
        & \colhead{R.A.}              %% R.A.         2
        & \colhead{Dec.}              %% Dec.         3
        & \colhead{$V_{\rm{lsr}}$}    %% V_lsr        4
        & \colhead{$\Delta V$}        %% del_v        5
        & \colhead{$S_{\rm{peak}}$}   %% S_peak       6
        & \colhead{$S_{\rm{int}}$}    %% S_int        7
        & \colhead{Distance}          %% dist         8
        & \colhead{Prob.}             %% prob of dist 9
        & \colhead{Code} \\           %% group       10 
        %% Unit
          \colhead{}                  %% Source Name
        & \colhead{(hh:mm:ss.sss)}    %% R.A.
        & \colhead{(dd:mm:ss.ss)}     %% Dec.
        & \colhead{(\kms)}            %% V_lsr
        & \colhead{(\kms)}            %% del_v
        & \colhead{(Jy beam$^{-1}$)}  %% S_peak
        & \colhead{(Jy)}              %% S_int
        & \colhead{(kpc)}             %% |
        & \colhead{}                  %% prob
        & \colhead{}                  %% dc_flag
      }
      \startdata
      G000.546$-$0.851&17:50:14.523&$-$28:54:31.25&  15.06&   8.95&   90.75& 1091.44&  2.7$\pm$0.3&0.59&A \\
      G000.644$-$0.042&17:47:18.663&$-$28:24:24.82&  49.95&   6.67&   55.06&  497.05& 10.8$\pm$0.2&0.54&B \\
      G000.647$-$0.055&17:47:22.052&$-$28:24:42.53&  51.17&   1.05&    3.95&   16.19& 10.8$\pm$0.2&0.58&B \\
      G000.651$-$0.048&17:47:21.125&$-$28:24:18.12&  47.92&   8.96&   15.45&   73.89&  8.3$\pm$0.3&0.54&N \\
      G000.656$-$0.040&17:47:20.062&$-$28:23:46.65&  52.42&   5.62&    6.47&   35.99& 10.8$\pm$0.2&0.63&A \\
      \enddata 
      \tablecomments{This is a sample of the 6.7~GHz methanol maser catalog.
                     Source names are based on the Galactic coordinates.
                     J2000 coordinates are presented in sexagesimal terms. 
                     $V_{\rm{lsr}}$ is the radial velocity of the strongest spot
                     of the source. $S_{\rm{peak}}$ and $S_{\rm{int}}$ are the
                     peak and integrated fluxes. Distances with uncertainties 
                     are presented in kpc. The probabilities evaluate how likely
                     the sources are located on the given distances. 
                     See Sec.~\ref{subsec:Prop_Distance} for details.
                     In the last column, code indicates 
                     the association between maser source and the continuum 
                     emission.  ``A'' means the associated continuum emission 
                     is detected and classified as group A, ``B'' means 
                     continuum emission is found in the vicinity of the maser 
                     but with no evidence supporting physical relationship 
                     between maser and continuum source. ``N'' represents the 
                     absence of continuum emission.  Only 5 entries of the 
                     catalog are shown here. For the entire machine-readable 
                     catalog, please see online material.
                    }
    \end{deluxetable}

  % spots catalogue
    
  \subsection{Methanol Maser Spot Catalog}
  \label{subsec:Produc_SpotCatalogue}
  
\begin{figure}
      %\resizebox{\hsize}{!}{\includegraphics[scale=0.5]{figures/spotmap_example.pdf}}
      %\plotone{figures/spotmap_example.pdf}
      \center
      \includegraphics[scale=0.7]{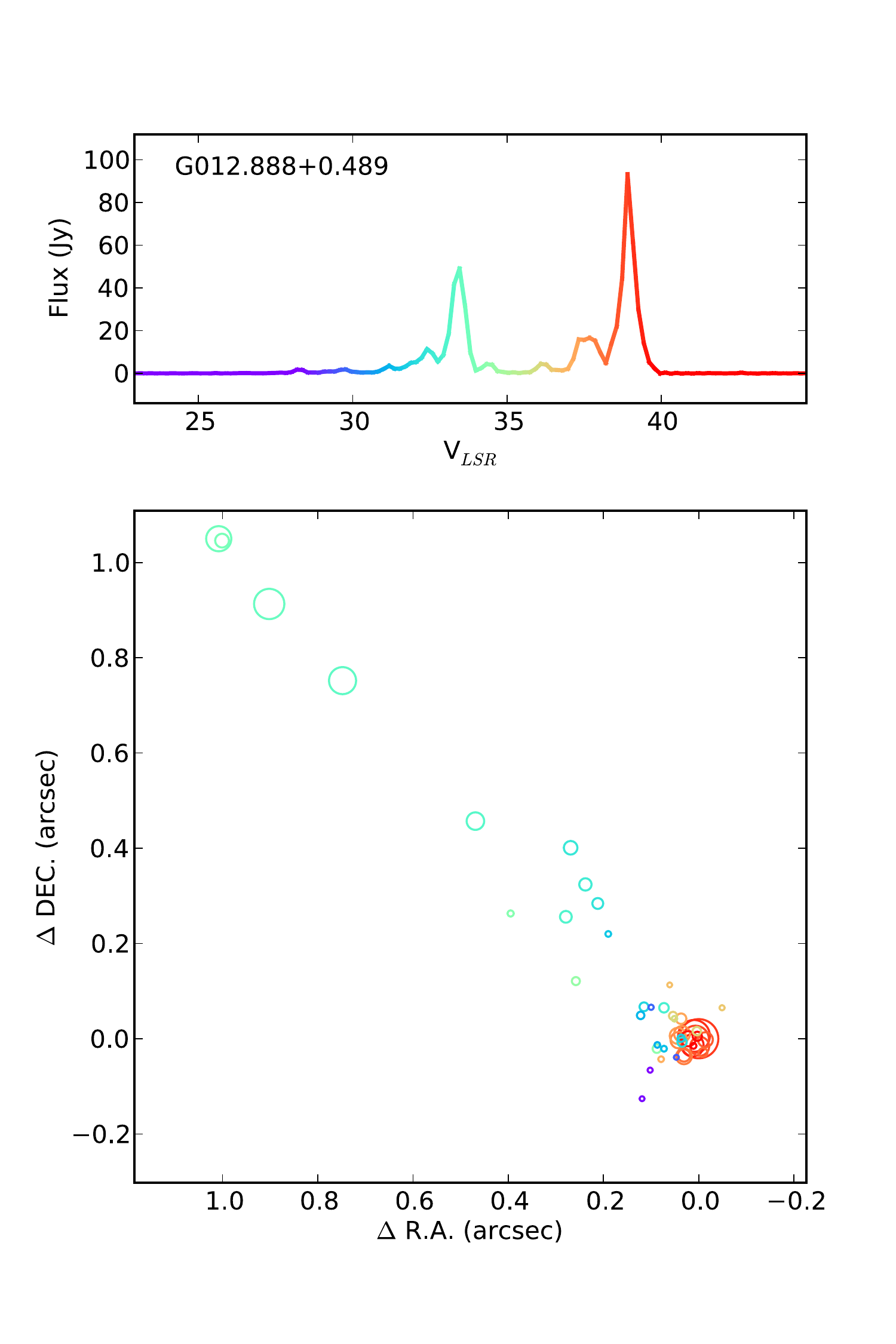}
      \caption{An example of the 6.7 GHz maser data for G012.888+0.489. 
               The {\it upper panel} shows the maser source spectrum and the
               {\it lower panel} shows a map of the maser spots.  Both plots have
               color-coded radial velocities.  The diameter of each spot is 
               proportional to the flux. 
               Positional offsets are relative to the strongest maser spot. 
               Spectra and maps for all sources are available as
               online material}
      \label{fig:Spotmap}
    \end{figure}

    Maser spots extracted from line cubes constitute a second catalog.  
    In order to facilitate
    the investigation in morphology and kinematics of individual maser source,
    we present maser spots arranged by the hosting maser source name in
    Table~\ref{tab:SpotCatalogue}.  In this spots catalog, columns in the 
    catalog are maser source name, reference position of each source, LSR 
    velocity, peak flux density, and the offsets of spot from the reference 
    position with their uncertainty. The reference position are determined
    conventionally close to the strongest maser spot in each source, thus bear
    no actually physical meaning. One should also notice that, in the catalog 
    the offsets of individual maser spot is provided uniformly by SAD task in 
    AIPS, which are listed to 3 decimals in arc second. However as the SNR of
    the single channel image varies from channel to channel and source to
    source, such a presentation may not indicate the actual precision of
    individual measurement.  The nominal precision of maser sources
    is $\approx27$~mas in the E-W direction and $\approx54$~mas in the N-S 
    direction (See Section~\ref{subsec:Produc_MaserCatalogue}).   
    From this catalog, we produced a spot map image and 
    a spectrum for each maser source.  See Figure~\ref{fig:Spotmap} for 
    an example. The full-size machine-readable spot catalog and images for 
    367 sources are available as online material.

    % spot catalogue
\begin{deluxetable}{rrrrrrrrr}
  \rotate
  \tablecolumns{9}
  \tablewidth{0pc}
  \tablecaption{The Maser Spots Catalog\label{tab:SpotCatalogue}}
  \tablehead{
    %% Header
      \colhead{Name}             %% Source Name   1
    & \colhead{Ref. R.A.}        %% Reference RA  2
    & \colhead{Ref. Dec.}        %% Reference Dec 3
    & \colhead{$V_{\rm{lsr}}$}  %% V_lsr         4
    & \colhead{$S_{\rm{peak}}$} %% S_peak        5
    & \colhead{$\delta X$}      %% del_x         6
    & \colhead{Unc.}             %%               7
    & \colhead{$\delta Y$}      %% del_y         8
    & \colhead{Unc.}    \\     %%               9
    %% Unit
      \colhead{}                 %% Source Name   1
    & \colhead{(hh:mm:ss.sss)}   %% Reference RA  2
    & \colhead{(dd:mm:ss.ss)}    %% Reference Dec 3
    & \colhead{(\kms)}          %% V_lsr         4
    & \colhead{(\Jypb)}         %% S_peak        5
    & \colhead{(\arcsec)}       %% del_x         6
    & \colhead{(\arcsec)}       %%               7
    & \colhead{(\arcsec)}       %% del_y         8
    & \colhead{(\arcsec)}       %%               9
  }
  \startdata
G012.888+0.489&18:11:51.392&$-$17:31:30.40&$+$38.91&   93.21&0.035&0.001&-0.80&0.001 \\
G012.888+0.489&18:11:51.392&$-$17:31:30.40&$+$39.08&   61.47&0.045&0.001&-0.80&0.002 \\
G012.888+0.489&18:11:51.392&$-$17:31:30.40&$+$33.46&   55.85&0.937&0.001&0.104&0.002 \\
G012.888+0.489&18:11:51.392&$-$17:31:30.40&$+$33.29&   44.61&0.783&0.001&-0.05&0.002 \\
G012.888+0.489&18:11:51.392&$-$17:31:30.40&$+$38.73&   44.40&0.042&0.001&-0.81&0.002 \\
G012.888+0.489&18:11:51.392&$-$17:31:30.40&$+$33.64&   38.48&1.043&0.001&0.241&0.002 \\
G012.888+0.489&18:11:51.392&$-$17:31:30.40&$+$39.26&   29.89&0.048&0.001&-0.82&0.002 \\
G012.888+0.489&18:11:51.392&$-$17:31:30.40&$+$38.56&   21.94&0.033&0.001&-0.82&0.003 \\
G012.888+0.489&18:11:51.392&$-$17:31:30.40&$+$33.11&   18.65&0.504&0.002&-0.35&0.004 \\
G012.888+0.489&18:11:51.392&$-$17:31:30.40&$+$37.68&   16.56&0.077&0.001&-0.81&0.004 \\
G012.888+0.489&18:11:51.392&$-$17:31:30.40&$+$37.33&   15.86&0.079&0.002&-0.80&0.004 \\
G012.888+0.489&18:11:51.392&$-$17:31:30.40&$+$37.50&   15.64&0.070&0.002&-0.79&0.004 \\
G012.888+0.489&18:11:51.392&$-$17:31:30.40&$+$37.85&   15.23&0.066&0.002&-0.84&0.004 \\
G012.888+0.489&18:11:51.392&$-$17:31:30.40&$+$39.44&   14.13&0.057&0.002&-0.80&0.004 \\
G012.888+0.489&18:11:51.392&$-$17:31:30.40&$+$38.38&   13.71&0.021&0.002&-0.81&0.004 \\
G012.888+0.489&18:11:51.392&$-$17:31:30.40&$+$33.82&   11.67&1.036&0.002&0.237&0.005 \\
G012.888+0.489&18:11:51.392&$-$17:31:30.40&$+$32.41&   11.33&0.304&0.002&-0.40&0.005 \\
G012.888+0.489&18:11:51.392&$-$17:31:30.40&$+$38.03&    9.62&0.066&0.002&-0.84&0.006 \\
G012.888+0.489&18:11:51.392&$-$17:31:30.40&$+$32.59&    9.32&0.273&0.003&-0.48&0.006 \\
G012.888+0.489&18:11:51.392&$-$17:31:30.40&$+$32.94&    8.50&0.314&0.003&-0.55&0.006 \\
G012.888+0.489&18:11:51.392&$-$17:31:30.40&$+$32.24&    7.24&0.247&0.003&-0.52&0.007 \\
G012.888+0.489&18:11:51.392&$-$17:31:30.40&$+$37.15&    6.79&0.072&0.003&-0.76&0.008 \\
G012.888+0.489&18:11:51.392&$-$17:31:30.40&$+$32.76&    5.54&0.108&0.004&-0.74&0.009 \\
G012.888+0.489&18:11:51.392&$-$17:31:30.40&$+$32.06&    5.33&0.070&0.004&-0.81&0.009 \\
G012.888+0.489&18:11:51.392&$-$17:31:30.40&$+$39.61&    5.21&0.038&0.004&-0.80&0.010 \\
G012.888+0.489&18:11:51.392&$-$17:31:30.40&$+$31.88&    4.83&0.150&0.004&-0.74&0.010 \\
G012.888+0.489&18:11:51.392&$-$17:31:30.40&$+$38.21&    4.83&0.026&0.005&-0.83&0.011 \\
G012.888+0.489&18:11:51.392&$-$17:31:30.40&$+$36.10&    4.41&0.039&0.005&-0.79&0.011 \\
G012.888+0.489&18:11:51.392&$-$17:31:30.40&$+$34.34&    4.34&0.123&0.005&-0.83&0.011 \\
G012.888+0.489&18:11:51.392&$-$17:31:30.40&$+$36.27&    4.04&0.089&0.006&-0.76&0.013 \\
G012.888+0.489&18:11:51.392&$-$17:31:30.40&$+$34.52&    3.98&0.293&0.006&-0.68&0.013 \\
G012.888+0.489&18:11:51.392&$-$17:31:30.40&$+$31.18&    3.46&0.157&0.006&-0.76&0.015 \\
G012.888+0.489&18:11:51.392&$-$17:31:30.40&$+$31.71&    3.19&0.072&0.006&-0.80&0.014 \\
G012.888+0.489&18:11:51.392&$-$17:31:30.40&$+$34.17&    2.52&0.430&0.008&-0.54&0.020 \\
G012.888+0.489&18:11:51.392&$-$17:31:30.40&$+$39.79&    2.24&0.046&0.009&-0.82&0.021 \\
G012.888+0.489&18:11:51.392&$-$17:31:30.40&$+$31.36&    2.20&0.108&0.009&-0.83&0.022 \\
G012.888+0.489&18:11:51.392&$-$17:31:30.40&$+$31.53&    2.17&0.225&0.010&-0.58&0.023 \\
G012.888+0.489&18:11:51.392&$-$17:31:30.40&$+$36.98&    2.12&0.114&0.010&-0.85&0.022 \\
G012.888+0.489&18:11:51.392&$-$17:31:30.40&$+$31.01&    2.07&0.122&0.009&-0.82&0.022 \\
G012.888+0.489&18:11:51.392&$-$17:31:30.40&$+$35.92&    2.00&0.086&0.012&-0.76&0.027 \\
G012.888+0.489&18:11:51.392&$-$17:31:30.40&$+$29.78&    1.81&0.135&0.012&-0.74&0.028 \\
G012.888+0.489&18:11:51.392&$-$17:31:30.40&$+$36.45&    1.71&-0.01&0.012&-0.74&0.028 \\
G012.888+0.489&18:11:51.392&$-$17:31:30.40&$+$28.20&    1.70&0.137&0.012&-0.87&0.029 \\
G012.888+0.489&18:11:51.392&$-$17:31:30.40&$+$29.60&    1.61&0.082&0.012&-0.84&0.030 \\
G012.888+0.489&18:11:51.392&$-$17:31:30.40&$+$28.37&    1.57&0.154&0.013&-0.93&0.029 \\
G012.888+0.489&18:11:51.392&$-$17:31:30.40&$+$36.63&    1.54&0.096&0.014&-0.69&0.033 \\
      \enddata
      \tablecomments{This table demonstrates the maser spots in source
                     G012.888+0.489 as an example of the spot catalogs. 
                     The name of the source and the reference point are given 
                     in the first three columns. In the body of the table,
                     each entry is representing one spot, given the order from
                     brightest to faintest. Radial velocities and peak fluxes
                     from Gaussian fitting result are given in the second and the
                     third column. Column 6 to 9 are the offsets of the spot
                     relative to the reference point in arc~second. 
                     The full-size methanol maser spots catalog for all sources 
                     will be available in machine readable format via 
                     online material.}
    \end{deluxetable}  

  \subsection{Continuum Catalog}
  \label{subsec:Produc_ContCatalogue}
    A catalog of radio continuum sources nearest the masers is presented
    in Table~\ref{tab:Cont}, including the Galactic 
    source name, J2000 coordinates of the peak emission, integrated flux 
    densities, the ratio of integrated flux to peak pixel value,
    ellipse parameters describing source's shape, the spatial offset between 
    peak of the continuum source and the corresponding maser, 
    the diameter of continuum source derived directly from the semi-major axis 
    length, and a "Group" code.  The precision of continuum peak position comes
    from the similar deduction we have made for the maser spots. As the typical
    SNR value of continuum images is 10, the positional precisions on E-W and
    N-S direction are $\approx135$ mas and $\approx270$ mas. The integrated 
    flux density~($F_{\rm{int}}$) 
    is derived from 2-D Gaussian fitting of the continuum brightness 
    distribution profile.  The ratio of integrated flux to peak pixel value 
    provides a measure of the compactness of the source, which is useful
    for morphological analysis of the continuum sources.  The ellipse parameters 
    $a$, $b$ and $\theta$ are the length of the major, minor axes
    and the positional angle counting East of North, respectively. 
    Elliptical parameters are derived from the 2-D Gaussian fitting that
    deconvolved from the restoring beam, if the source is resolved on
    both major and minor axis directions, otherwise, parameters are of
    the 2-D Gaussian component that convolved with the restoring beam.
    We regard them as estimates of source's angular size, and the parameter 
    $a$ is used to calculate a spatial diameter using the heliocentric distance 
    of the associated maser.  However, our observations are lacking 
    sufficient resolving power to measure the actual size of some extremely 
    compact and distant sources.  We indicate unresolved sources with a ``$<$''
    symbol when the formal fitted FWHM was smaller than the CLEAN restoring 
    beam in any direction.

    We divide continuum sources into two groups:
    Group A includes 140 continuum sources that are 
    associated with a methanol maser, whereas
    139 Group B sources have no associated maser sources. 
    See Sec~\ref{subsec:Prop_Association} for more detailed discussion on 
    this issue.

    % continuum catalogue
    \begin{deluxetable}{rrrrrrrrrrr}
      \rotate
      \tablecolumns{11}
      \tablewidth{0pc}
      \tablecaption{Catalog of radio continuum counterpart of maser sources
                    \label{tab:Cont}}
      \tablehead{
      %% Header
       \colhead{Source Name}
      &\colhead{R.A.}
      &\colhead{Dec.}
      &\colhead{$F_{\rm{int}}$}
      &\colhead{$I/P$}
      &\colhead{$a$}
      &\colhead{$b$}
      &\colhead{P.A.}
      &\colhead{offset}
      &\colhead{diameter}
      &\colhead{Group} \\
      %% Unit
       \colhead{}                                 %% Source Name
      &\colhead{(hh:mm:ss.ss)}                    %% R.A.
      &\colhead{(dd:mm:ss.s)}                     %% Dec.
      &\colhead{(mJy)}
      &\colhead{}
      &\colhead{(\arcsec)}
      &\colhead{(\arcsec)}
      &\colhead{(\degr\ )}
      &\colhead{(pc)}
      &\colhead{(pc)}
      &\colhead{} 
      }
      \startdata
      G001.328+0.149 & 17:48:09.95 & -27:43:09.0 &    138.00 &     59.87 & 44.8 & 41.7 &   1.9 & 0.27 &    0.91 &A\\
      G002.143+0.009 & 17:50:36.00 & -27:05:47.4 &      5.87 &      1.12 &  2.4 &  1.2 &  14.0 & 0.06 &    0.12 &A\\
      G002.535+0.198 & 17:50:51.30 & -26:39:52.5 &      3.79 &      2.59 &  5.8 &  4.1 &  73.0 & 1.34 &    0.12 &B\\
      G003.910+0.001 & 17:54:38.77 & -25:34:45.1 &     49.70 &      1.03 &  8.4 &  2.9 & 174.4 & 0.02 & $<$0.45 &A\\
      G004.393+0.078 & 17:55:21.59 & -25:06:12.5 &    124.00 &     28.42 & 52.3 & 26.9 & 139.4 & 6.38 &    3.71 &B\\
      \enddata
      \tablecomments{The names of the corresponding maser sources are listed in
                     the first column. Given, from the second to fourth column 
                     are J2000 coordinates, the integrated flux of the entire 
                     continuum source $F_{\rm int}$ and $I/P$, the ratio of 
                     integrated flux to peak pixel value. The shape of the 
                     source is represented by elliptic parameters $a$, $b$ and 
                     position angle.  Offset indicates the displacement between 
                     peaks of maser and continuum sources in pc.  Diameter 
                     column indicates the diameter of the continuum source 
                     derived from the major
                     axis $a$ and the heliocentric distance of the associated
                     maser. Notice that the values in this column with ``$<$'' 
                     represent the upper limit of the 
                     physical diameter of the corresponding continuum source, 
                     due to our limited resolving power. The last column implies
                     the association type of the continuum source. 
                     Here only 5 entries are given. The 
                     full catalog in machine-readable format is available via
                     online material.}
    \end{deluxetable}

\begin{figure} 
      \includegraphics[scale=0.7]{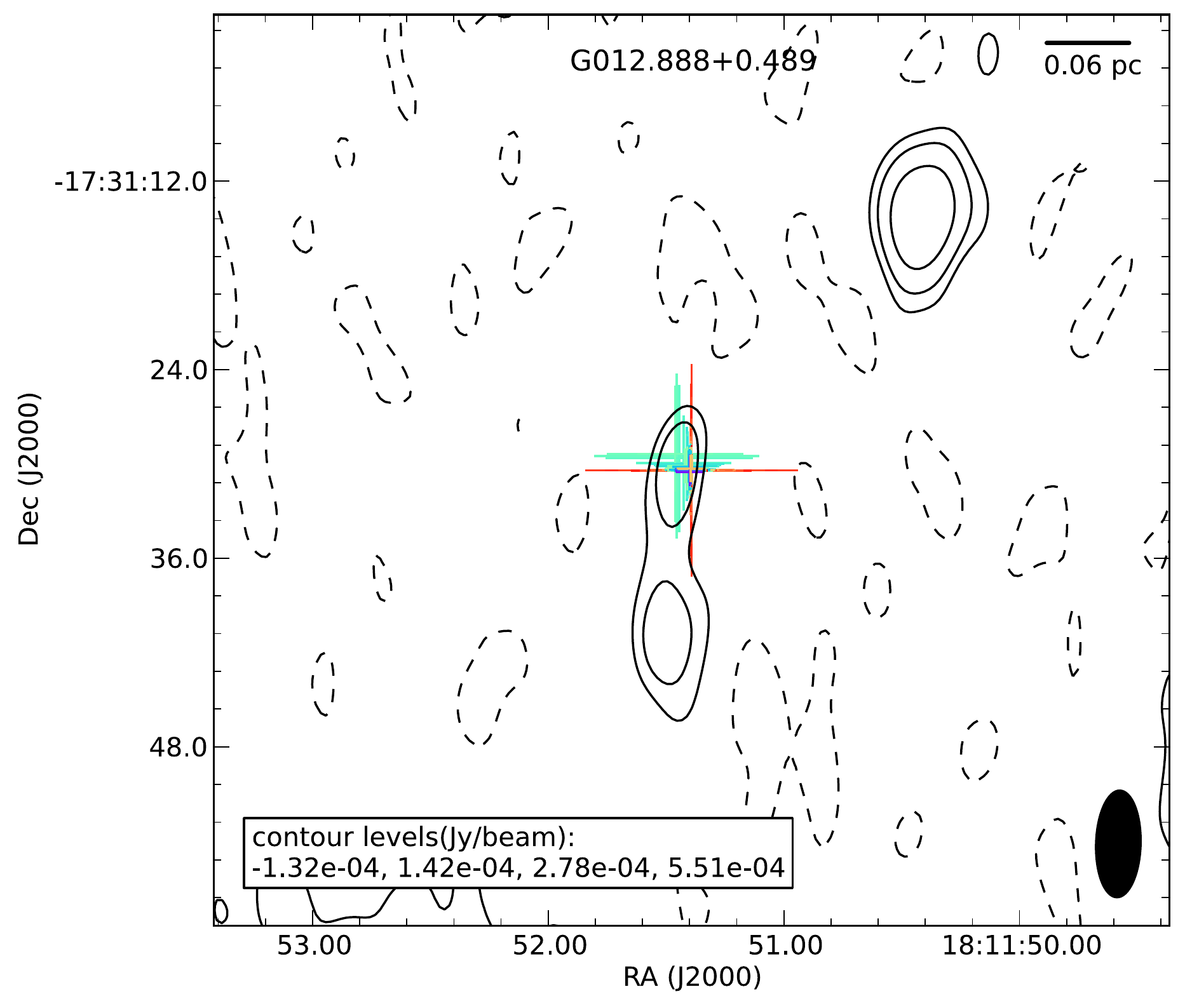} 
      \caption{An example radio continuum map ({\it contours}) with maser spots 
               ({\it crosses}) overlaid for G012.888+0.489
               Colors of the crosses indicate LSR velocity, as indicated in the
               source spectrum in Fig. \ref{fig:Spotmap}.
               For strong sources with peak brightness greater than $8\sigma$, 
               the contour levels are logarithmic at $2\sigma$, 
               $4\sigma$, $8\sigma$, $16\sigma$, etc.  For fainter sources the 
               contour levels are $2\sigma$, $4\sigma$, $6\sigma$, $8\sigma$, 
               etc.  Negative map brightnesses are indicated with dashed 
               contours.  The FWHM of the restoring beam is indicated by the 
               black ellipse at lower right corner.  A linear length scale 
               based on the estimated heliocentric distance of the maser is 
               shown in upper right corner.
               \label{fig:Overlap}}
    \end{figure}

    We overlaid the maser spots on the integrated continuum contour map to
    give a direct illustration of the positional relation of these two 
    kinds of detections. These images can be found 
    in the online material; one example is shown in Figure~\ref{fig:Overlap}
    in which the colored crosses are the maser spots and the black contours 
    indicate the radio continuum emission. 
    
\section{Source Properties}
\label{sec:Prop}

  \subsection{Positions}
  \label{subsec:Prop_Positions}
  
\begin{figure}
      \includegraphics[scale=0.7]{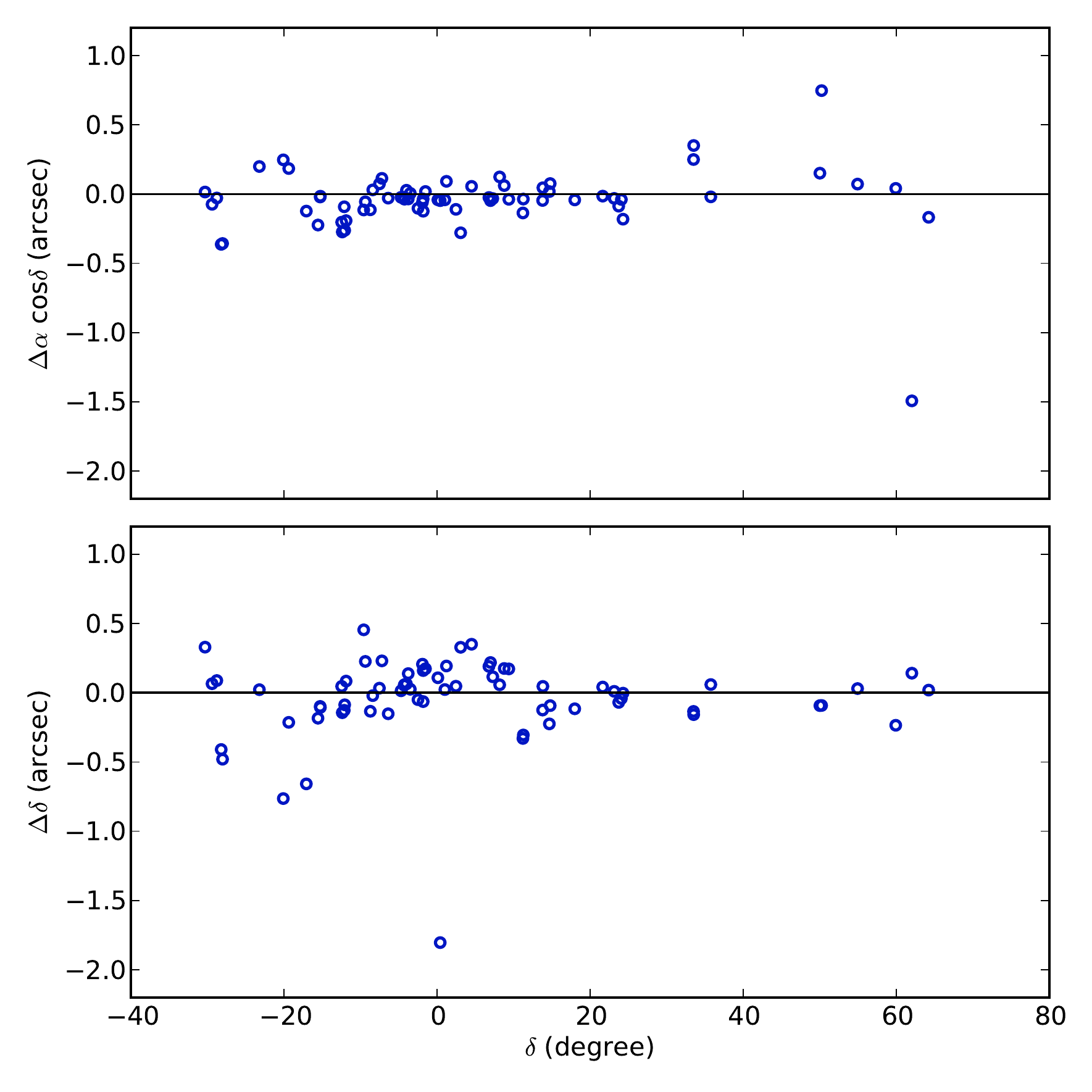}
      \caption{Position error as a function of source declination, obtained by   
               subtracting high accuracy BeSSeL Survey VLBI positions from our 
               VLA maser positions.  {\it Upper panel:} 
               E-W~($\Delta\alpha\cos\delta$) errors with an RMS 0.23\arcsec;
               they show no significant trend with source declination (beyond 
               2 outliers).  {\it Lower panel:} N-S~($\Delta\delta$) errors 
               with an RMS of 0.30\arcsec; they show a possible increase 
               below a Decl. of $-$10\degr. 
              }
      \label{fig:PositionUnc}
    \end{figure}

    The absolute positions of masers come from the images without 
    self-calibration.  We estimate position uncertainty by comparing
    our result with the VLBI measurement form the first epoch result of
    BeSSeL Survey\footnote{\url{http://bessel.vlbi-astrometry.org/first_epoch}}
    observed by VLBA and having better than 10 mas accuracy. 
    Fig.~\ref{fig:PositionUnc} shows the differences between the sources in 
    common to the two surveys, which we attribute to errors in the VLA data.
    In E-W direction, the RMS errors are about 0.23\arcsec, and in the N-S 
    direction they are about 0.30\arcsec, and somewhat larger for
    $\delta < -10\degr$.  The positional uncertainty of masers is mainly 
    limited by atmospheric irregularities that are only partially canceled by
    the small separation ($<5$\degr) between phase calibrators and targets. 
    
  \subsection{Distance}
  \label{subsec:Prop_Distance}
  
    \citet{2014ApJ...783..130R} have published a summary of the parallax 
    measurements of masers that are hosted by Galactic MSFRs.  For a total of 
    53 of our sources parallax measurements do exist.  While, in principle 
    kinematic distance can be calculated for sources in the Galactic plane, 
    these can be very inaccurate in the general direction of the Galactic 
    center and anti-center directions and within the Solar circle they are 
    multi-valued, i.e., are affected by the ``kinematic distance ambiguity'', 
    KDA.  In addition, kinematic distance estimates depend on the assumed 
    Galactic rotation curve, as well as the assumption of circular Galactic 
    orbits.

    \citet{2016ApJ...823...77R} proposed an alternative approach that 
    leverages VLBI parallax measurements of massive star forming regions 
    from the BeSSeL Survey
    and the Japanese VERA project to significantly improve the accuracy and
    reliability of distance estimates to any source known to
    follow spiral structure.  Their Bayesian approach considers the likelihood 
    of a spiral arm assignment, the angular displacement from the Galactic 
    plane, the proximity to individual sources with parallax measurements, 
    as well as kinematic information to estimate a full distance probability
    density function for each source.  Since this method not only synthesizes 
    different approaches of distance determination, but also shows good 
    agreement with results from the \ion{H}{1} absorption 
    method when resolving the KDA and can be applied to the full Galactic 
    longitude range, we adopt it to estimate distances (and uncertainties) 
    to the 367 methanol maser sources in our catalog, based on the radio
    velocity of the reference spot in each source. However, one should
    notice that the determination of the distances strongly depends on the
    spiral arm assignment. We simply take the association with the highest
    probability, so the distances of some sources may be more uncertain than
    what is claimed by the nominal uncertainty. In the methanol maser catalog
    the probability of the most likely arm assignment is given. We suggest 
    that one should take special care when dealing with those low probability
    distances and their subsequent parameters, such
    as luminosities, spatial sizes, etc. 
 
\begin{figure}
      \includegraphics{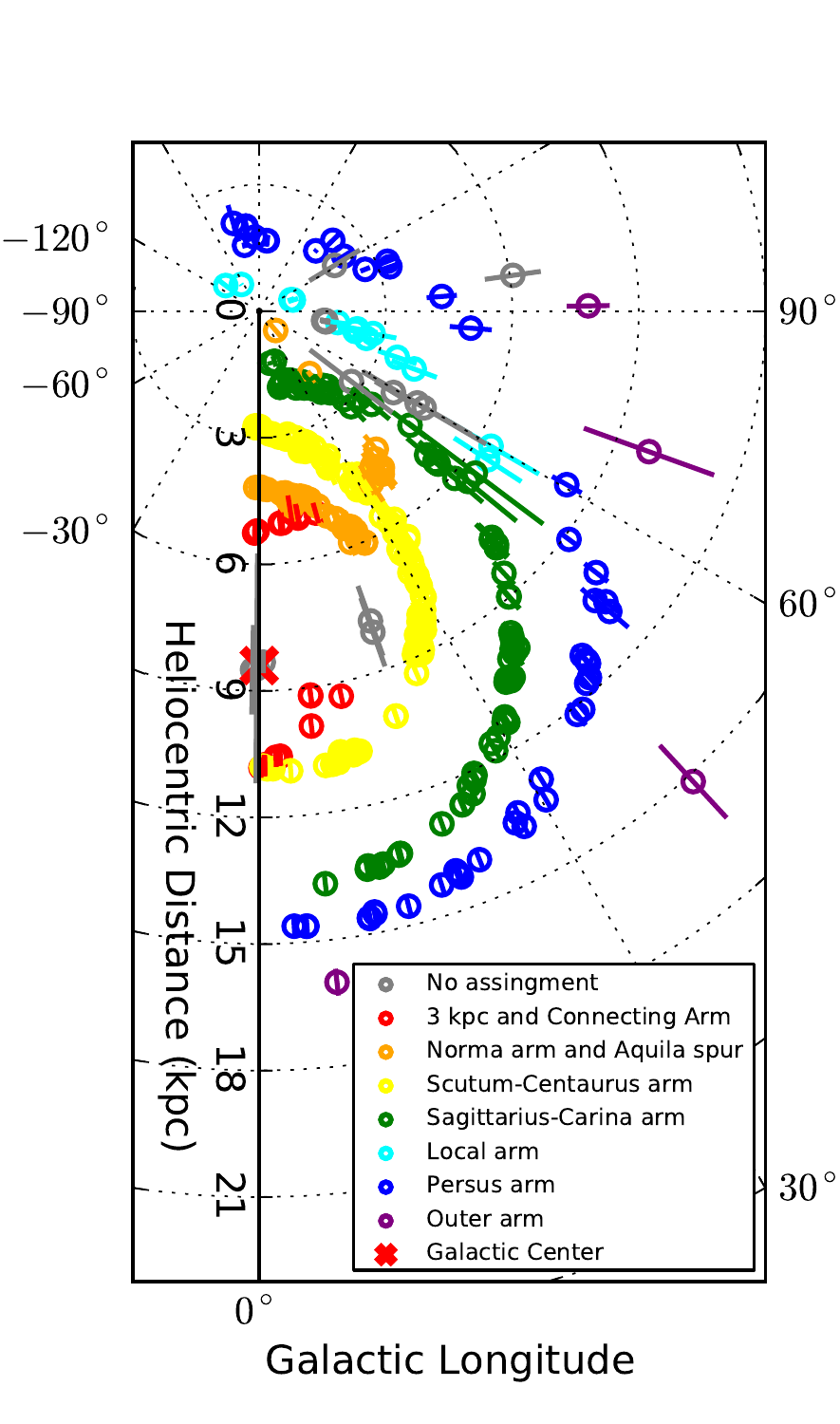}
      \caption{The Galactic plane distribution of methanol masers, with (color 
               coded) spiral arm associations and distances estimated by the 
               Bayesian approach of \citet{2016ApJ...823...77R}.  Error bars 
               indicated 
               the formal uncertainty of the distance estimates.  Grey circles 
               indicate the small number of sources for which no significant 
               arm assignment could be made. 
              }
      \label{fig:Gala}
    \end{figure}

    In Figure~\ref{fig:Gala},
    we show the distribution of all maser sources projected on the Galactic 
    plane.  Note, however, that the spiral arm models used to partially 
    constrain distances are assumed to be correct; the accuracy of these 
    models is limited to regions where parallaxes have been 
    measured~\citep[see][for the details]{2016ApJ...823...77R}.  

  \subsection{Association of Masers and Continuum sources}
  \label{subsec:Prop_Association}

\begin{figure}
      \includegraphics[scale=0.7]{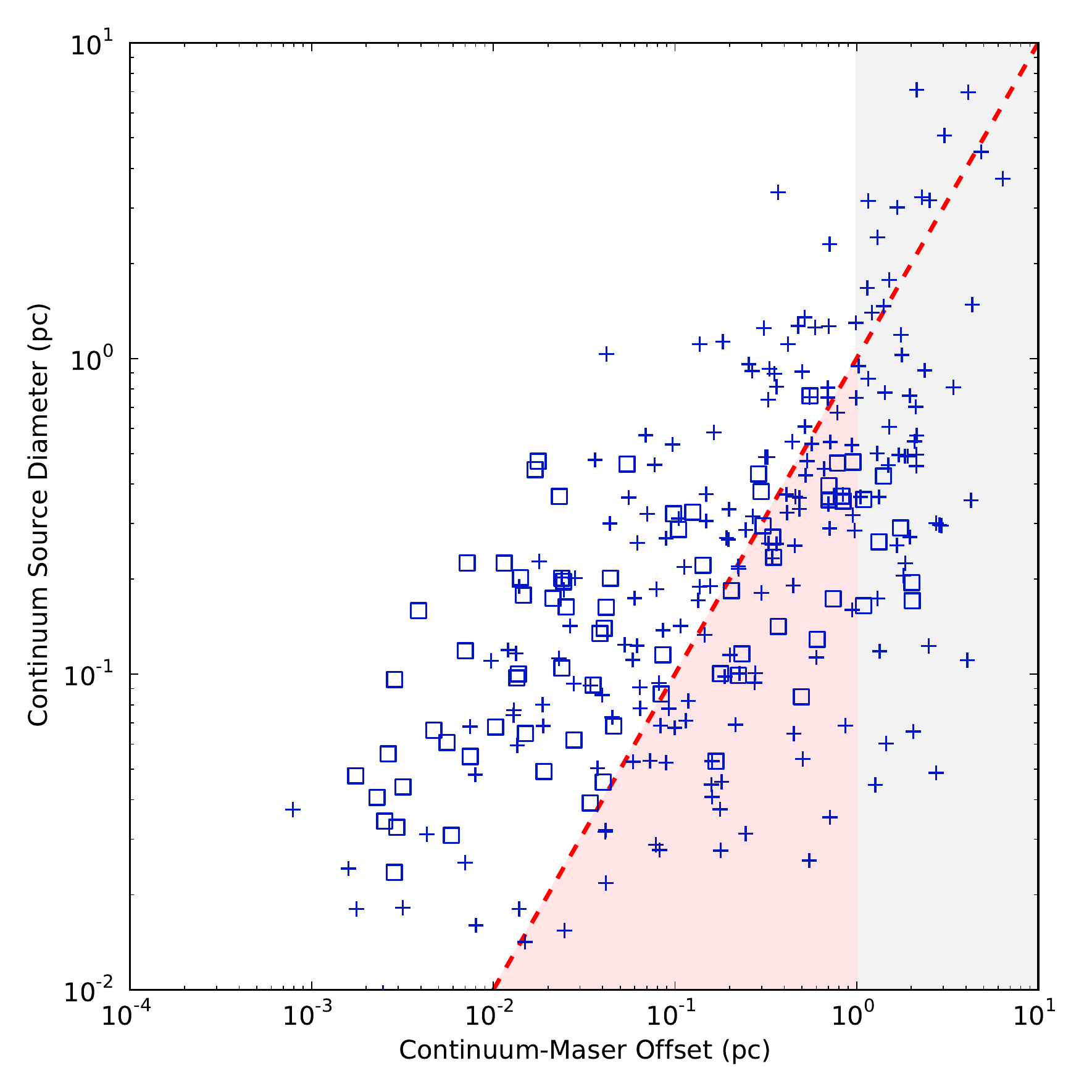}
      \caption{
              Continuum source diameter versus the offset between peak 
              positions of the continuum and reference maser spot emission.
              {\it Blue crosses} indicate continuum sources with size 
              measurements, while {\it blue squares} indicate sources with 
              only upper limits on size.  The {\it red dashed} line traces 
              where the continuum source size equals its offset from the maser.
              The {\it gray shaded} region indicates separations $>1$ pc, 
              which likely indicates no direct physical association.  Sources 
              in white area comprise Group A (likely associated), and those in 
              pink and gray area comprise Group B (not likely associated).
              }
      \label{fig:Association}
    \end{figure}

    We establish an association between a maser and a continuum source based
    on their projected separations and the size of the continuum source.  
    Since the two well accepted pumping mechanisms of the class II methanol 
    maser are infrared radiation from the central 
    YSO~\citep{1994A&A...291..569S} and collision~\citep{2005MNRAS.360..533C}, 
    the continuum-associated class II methanol
    masers can only exist in the natal molecular cloud of the HC/UC \ion{H}{2}
    region. Projected on the sky, the associated maser should appear within
    or in the vicinity of the area of the continuum source.      
    Thus, we introduce two criteria on the projected spatial distance 
    between peak positions of maser and continuum source when considering their
    association. First, since the median value of the radii of gas clumps 
    hosting methanol masers is 0.97~pc~\citep{2013MNRAS.435..400U}, 
    we assume the maser source is not physically associated with the continuum,
    if their projected offset is greater than 1~pc. Second,
    two objects are considered associated with each other only when the offset
    is smaller than the semi major axis of the continuum source.
    In Fig~\ref{fig:Association}, we plot the continuum source diameter versus 
    the offset to the maser reference spot.  The gray area in the plot 
    represents the offset larger than 1~pc, while the red dashed-line indicates 
    that the continuum source diameter equals the projected offset to the 
    maser.  The markers in gray area are denoting the maser-continuum pairs
    that do not in the same molecular cloud, and the those in the pink area
    can be interpreted as  the continuum source and the methanol maser trace 
    different massive star formation sites in the same molecular cloud. 
    In other words, maser-continuum pair in these two areas show no possibility
    of direct physical association.  We refer sources in these two area as 
    Group B. In total, 139 sources are found in this group.
    In contrast, we assign 140 continuum sources in the white area of the plot  
    (i.e. offsets less than either 1~pc or the diameter of continuum 
    source) as physically associated with their counterpart masers and
    refer to these as Group A.  Here we provide a rather rough classification.
    A more comprehensive diagnose on the association is still in need in the 
    following work, since the relationship of these two signposts will do
    great importance in the study of the evolutionary stages of HMSFRs.

  \subsection{Flux Relationship}
  \label{subsec:Prop_Flux}

\begin{figure}
      \includegraphics{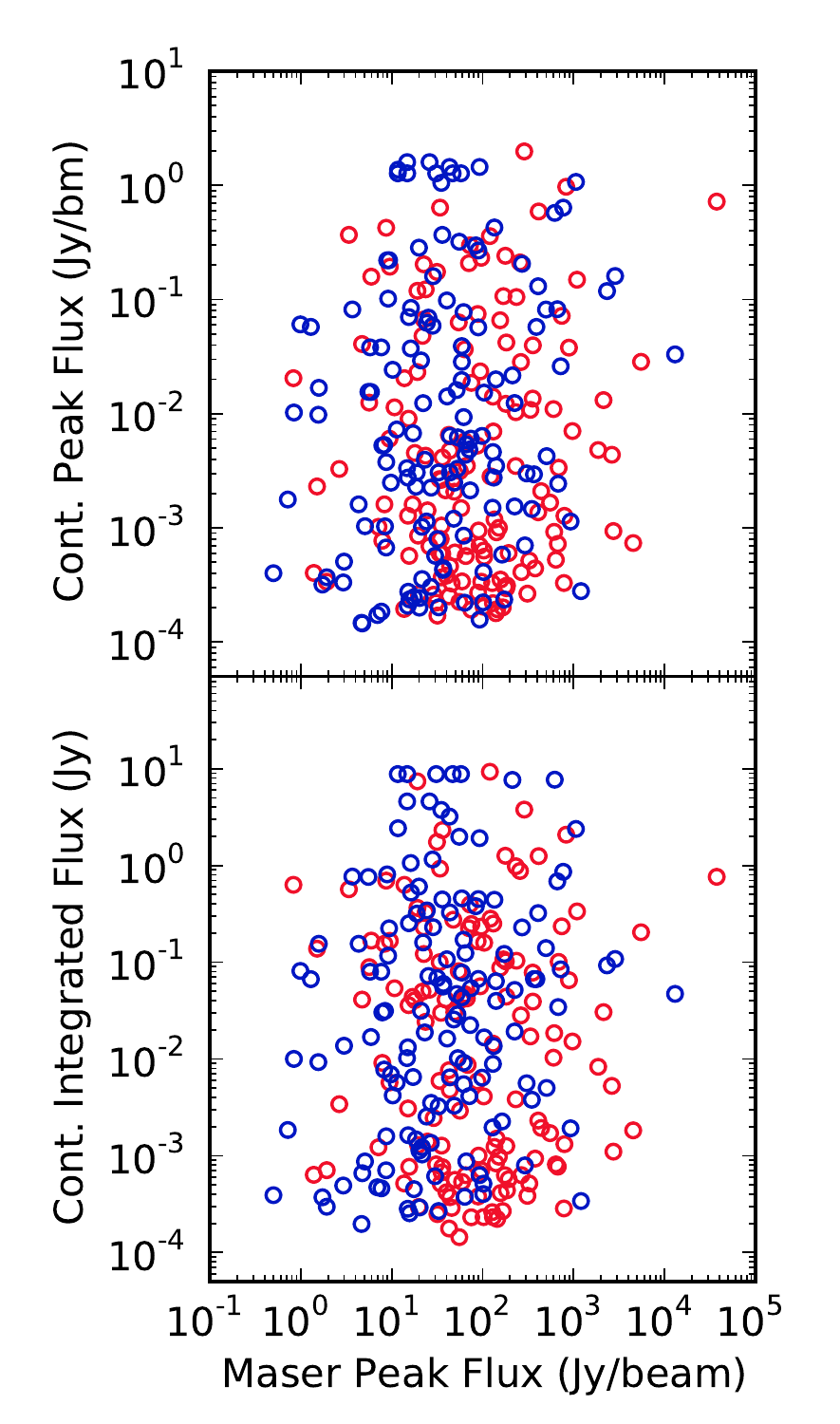}
      \caption{Continuum peak and integrated flux density plotted as function of
               maser peak flux density.  Sources in Group A and B are denoted
               by {\it red} and {\it blue circles}, respectively.  No relation 
               between continuum and maser strengths are apparent.}
      \label{fig:Flux}
    \end{figure}

    Fig~\ref{fig:Flux} shows the plots of continuum source peak and 
    integrated flux as function of maser peak flux.  Within our sample, we 
    found no obvious relationship between flux of methanol masers and 
    continuum sources in both group A and group B.  Thus we conclude that 
    methanol maser luminosity is independent from the luminosity of the 
    associated radio continuum source.

\section{Summary}
\label{sec:Summary}

  We report a survey toward a large sample of Galactic class II
  6.7~GHz methanol masers and their C-band radio continuum counterparts.
  With the VLA in its C-configuration, we observed 372 targets selected from 
  a variety of publications and produced a catalog of 367 methanol 
  masers with sub arc~second level positional accuracy, and high resolution
  radial velocity.  We believe our catalog is a uniform and comprehensive 
  collection of class II methanol maser in the first and second Galactic 
  quadrants. High accuracy J2000 coordinates, radial velocities 
  and fluxes are presented. 
  Maser spots detected for each source are reported in 
  367 individual on-line available files, and we plot spot maps and spectra
  for each maser source.
  We also search for compact radio continuum emission toward all masers.
  This resulted in detections of 279 radio continuum sources, of which 157 are 
  likely to be physically associated with a maser source.   
  Statistical characteristics of various source properties are presented.
  These three catalogs may aid future investigations of massive star forming 
  and the maser process.
  
\acknowledgments

  This work was supported by the National Natural Science Foundation of China 
  (Grant Number: 11133008).

% for the bibliography, at the end
\bibliographystyle{aasjournal} % style aasjournal.bst
% your references Yourfile.bib
\bibliography{./aasbib} 

\end{document}